\documentclass[a4paper]{jpconf}
\usepackage{graphicx}
\usepackage{iopams}

\begin{document}

\voffset-0.5in

\title{The Direction of Gravity} 

\author{Eric V.\ Linder}

\address{Berkeley Lab \& University of California, Berkeley, CA 94720 USA\\ 
Institute for the Early Universe WCU, Ewha Womans University, Seoul, Korea} 

\ead{evlinder@lbl.gov} 

\begin{abstract} 
Gravity directs the paths of light rays and the growth of structure.  
Moreover, gravity on cosmological scales does not simply point down: it 
accelerates the universal expansion by pulling outward, either due to a 
highly negative pressure dark energy or an extension of general relativity. 
We examine methods to test the properties of gravity through cosmological 
measurements. We then consider specific possibilities for a sound 
gravitational theory based on the 
Galileon shift symmetry.  The evolution of the laws of gravity from the 
early universe to the present acceleration to the future fate -- the paths 
of gravity -- carries rich information on this fundamental force of physics 
and on the mystery of dark energy. 
\end{abstract}

%%%%%%%%%%%%%%%%%%%%%%%%%%%%%%%%%%%%%%%%%%%%%%%%%%%%%%%%%%%% 
\section{Introduction} 

The direction of gravity, in both senses of the word, is a key characteristic 
of the universe.  The revolution of the late 1990s was the discovery 
that gravity did not exclusively pull down in the conventional, attractive 
way but that cosmic expansion showed an acceleration equivalent to gravity 
pulling outward.  Secondly, gravitation plays the major role in directing 
the contents of the universe, bending the paths of light and governing the 
growth of large scale structure.  

A new frontier has opened in exploring the force of gravity, understanding 
not only its spatial direction but its direction of energy and mass in the 
sense of a conductor directing an orchestra.  Some outstanding questions now 
being addressed include: Why is gravity pulling outward -- due to an unknown 
dark energy component or a new law?  Is gravity (e.g.\ Newton's constant) 
constant, or strengthening or weakening with time?  Does gravity govern 
the growth of cosmic structure in exactly the manner it governs cosmic 
expansion, as in the laws of general relativity, or are there 
further degrees of freedom?  Does gravity behave the same on all scales? 

At its most basic, we should test our understanding of cosmic gravity 
because we can, as a reality check.  Observations of the deflection of 
light by gravity (lensing) and the growth of matter clustering by gravity 
are reaching the stage where they can provide incisive information.  
Moreover, we should justify the long extrapolation of general relativity 
from the small scales where it is tested to the cosmic scales where it 
is applied (more than 
a factor of billion in length), and from high curvature to low curvature 
regions.  Tests of cosmic gravity are further motivated by the fact that 
the first two precision observational tests have basically shown gaping 
holes in our understanding: 
general relativity plus attractive matter fails to predict acceleration in 
the cosmic expansion, and general relativity plus attractive matter fails 
to agree with the growth and clustering of large scale structure. 

In Sec.~\ref{sec:indep} we explore methods for testing gravity in 
phenomenological and model independent approaches.  Adopting a specific 
model showing desirable characteristics, Galileon gravity, in 
Sec.~\ref{sec:gal} we illustrate the theoretical constraints an alternate 
theory of gravity must satisfy and the handles by which future data will 
be able to test stringently the nature of gravity.

%%%%%%%%%%%%%%%%%%%%%%%%%%%%%%%%%%%%%%%%%%%%%%%%%%%%%%%%%%%%%%%%%%%%%% 
\section{The Differences of Gravity \label{sec:indep}} 

To test general relativity one can either simply look for consistency 
with observations or one can measure parameters or quantities that would 
differ from the general relativity prediction if deviations exist.  
For an early work addressing some implications of deviations, e.g.\ 
from the inverse square law of attraction, see \cite{peebles02}.  We 
will concentrate here on cosmic scales, those above galaxy cluster 
lengths but well within the horizon, 5--500 Mpc. 

General relativity says that effects on the cosmic expansion are 
mirrored exactly in the growth of structure, with no extra gravitational 
degrees of freedom to cause an offset.  Therefore comparing the expansion 
history to the growth history is one of the major tests of the physics. 
This implies that it is crucial to fit for the expansion and growth 
simultaneously, to be alerted to any deviations.  Should the gravity 
theory be assumed (e.g.\ as general relativity) incorrectly, this will 
bias the fits of both the expansion and cosmological model.  The converse is 
true as well (e.g.\ assuming $\Lambda$CDM).  Thus, even if one is not 
interested in gravity one still must fit for it or the quantities one is 
interested in will be biased. 

To separate cleanly the gravity effects on growth from the expansion 
effects on growth, the gravitational growth index parameter $\gamma$ was 
developed by \cite{lin05}.  Recall that linear growth of a fractional 
density perturbation $\delta\rho/\rho$ as a function of expansion factor 
$a$ can be well approximated by 
\begin{eqnarray} 
\frac{\delta\rho}{\rho}(a)&=&\frac{\delta\rho}{\rho}(a_i) \times 
e^{\int_{a_i}^a (da'/a')\,\Omega_m^\gamma(a')} \\ 
\Omega_m(a)&=&\frac{\Omega_m a^{-3}}{\Omega_m a^{-3}+(1-\Omega_m) 
e^{3\int_a^1 (da'/a')\,[1+w(a')]}} \ , 
\end{eqnarray} 
where $\Omega_m$ is the present matter density in units of the critical 
density and the dark energy equation of state $w(a)=w_0+w_a(1-a)$ to 
an excellent approximation.  This accurately deconvolves the two influences 
on the growth, the expansion history in the form of $\Omega_m(a)$ from the 
gravity law in the form of $\gamma$, providing 
a fit to better than 0.1\% in many of the standard cases.  Suppose one 
considered growth measurements arising from weak gravitational lensing 
surveys.  The weak lensing signal involves a convolution of the gravitational 
growth of massive structures at various redshifts along the line of sight 
and the geometric distance factors serving as a focal length.  If one 
assumes general relativity (GR), omitting to fit for gravity, then the dark 
energy parameters one derives from the (GR) growth and distances could be 
strongly biased.  The fit for the time variation of the dark energy equation 
of state $w_a$ will be incorrect by $\Delta w_a\approx 8\Delta\gamma$, where 
$\Delta\gamma$ is the amount by which the true gravity deviates from GR. 
That is, failing to fit for gravity also induces failure to fit expansion 
accurately.  The converse is true as well.  
Fitting for gravity is therefore necessary. 

If we want to go beyond the coarse grained approach of $\gamma$, how should 
we do so in a practical manner?  Adding parameters without careful thought 
will merely 
blow up the uncertainties in both gravity and expansion.  As with the 
gravitational growth index $\gamma$, one must carefully study how to 
separate cleanly the different physical influences.  One could 
work within a specific model that determines the gravity and expansion 
behaviors but such one by one comparisons are time consuming and often 
give little general insight.  Instead we consider various approaches to 
a more model independent analysis. 

General relativity, as summarized by John Wheeler, is that ``matter tells 
spacetime how to curve, and spacetime tells matter how to move''.  These 
statements can be thought of mathematically as Poisson's equation 
$\nabla^2\phi=4\pi G_N a^2 \delta\rho$ and Newton's first law 
$-\vec\nabla\psi=\ddot x$.  We have suggestively written the spacetime 
potentials with two different symbols: $\phi$ and $\psi$.  In GR they 
are one and the same, but in other theories of gravity they can differ.  
To tie them closely to the observations we consider two modified 
Poisson equations, 
\begin{eqnarray} 
\nabla^2(\phi+\psi)&=&8\pi G_N a^2\delta\rho \times G_{\rm light}\\ 
\nabla^2 \psi&=&4\pi G_N a^2\delta\rho \times G_{\rm matter} \ . 
\end{eqnarray} 
The function $G_{\rm light}$ tests how light responds to gravity, and 
is central to gravitational lensing and integrated Sachs-Wolfe measurements. 
The function $G_{\rm matter}$ tests how matter responds to gravity, and 
is central to growth of massive structures and peculiar velocities; the 
growth index $\gamma$ is closely related.  Also see \cite{edbert} for a 
more detailed discussion of these modified Poisson equations. 

In general $G_{\rm light}$ and $G_{\rm matter}$ will be functions of 
time and space, e.g.\ redshift $z=a^{-1}-1$ and length scale $r$ or 
wavemode $k$. 
One approach to parametrizing them with a practical number of degrees of 
freedom is to be guided by classes of gravity theories.  Looking at both 
higher dimension gravity and scalar-tensor theories one sees a deviation 
starting from early universe GR behavior and involving either a scale 
independent or characteristic $k^2$ dependence (arising from the Laplacian 
in the equations of motion).  Thus, a useful form is \cite{zhao1109} 
\begin{equation} 
G_{\rm matter}=1+\frac{ca^s (k/H_0)^n}{1+3|c|a^s (k/H_0)^n} \ , \label{eq:pade} 
\end{equation} 
where $n=0$, 2 respectively and a similar form could hold for $G_{\rm light}$. 
Note that leaving out the denominator, so $G$ simply deviates from the GR 
value of 1 as $a^s$ is dangerous as it unfairly weights high vs low redshift 
observations and can easily lead to bias.  The Pad{\'e} form of 
Eq.~(\ref{eq:pade}) is accurate to $\sim$1\% for DGP and $f(R)$ gravity. 

For current data, involving weak gravitational lensing, galaxy peculiar 
velocities, CMB power spectra, supernova distances, and baryon acoustic 
oscillations, Ref.~\cite{zhao1109} simultaneously fits for gravity (the 
amplitude 
$c$ and time dependence $s$) and expansion (matter density $\Omega_m$ and 
dark energy equation of state $w_0$, $w_a$).  They find the gravity 
constraints are 
nearly unaffected by the expansion fit, and the expansion constraints are 
nearly unaffected by the gravity fit.  To a significant extent, however, 
this reflects the status of current data: almost no constraint exists on 
$s$.  As data improves and becomes more incisive in testing gravity we do 
expect some degradation in expansion constraints when also fitting, vs 
fixing, gravity -- perhaps a factor of $\sim2$ in confidence contour area. 

Both the above theories of gravity mentioned are simple models with 
restricted parameters.  DGP \cite{dgp} has one gravity parameter, the 
5D crossover scale that determines both the amplitude and time dependence 
of the deviations from GR.  The $f(R)$ family of scalar-tensor models 
has more freedom but can successfully be treated in terms of the 
scalaron mass \cite{scalaron}, and this is well fit by an amplitude 
(of order $\sim100 H_0$) and a power law time dependence (i.e.\ $s$) -- 
basically two parameters. 

Indeed, if one considers the phase space, or ``paths of gravity'', of 
the deviations from GR, DGP shows a definite, near parabolic track in 
the $G'_{\rm matter}$--$G_{\rm matter}$ plane and the family of $f(R)$ 
tracks can be calibrated through a rescaling of $s$ into a single, also 
near parabolic track \cite{roysoc}.  See Figure~3 of \cite{roysoc} for 
an illustration and further details.  These two classes diverge from each 
other, and GR, having today $\Delta G_{\rm matter}\approx \pm 0.3$.  This 
suggests, in these cases at least, that we should strive for observations 
capable of testing this beyond Einstein parameter at the 10\% level or 
better. 

Not every viable theory of gravity may be so simple, however.  It is 
useful to have another, more model independent approach in our toolkit. 
A reasonable choice is to test for the values of the $G$ functions in bins 
of redshift and wavenumber, i.e.\ asking whether gravity behaves the same 
(and like GR) at high and low redshift, and for high and low wavenumbers 
(small and large scales).  While basis functions or principal components 
are other possible choices, even next generation data will not have the 
leverage to fit more than 2 modes each with reasonable signal to noise 
(not just low noise).  Recall that $N$ modes in each of redshift and 
wavenumber gives rise to $2N^2$ parameters (for $G_{\rm matter}$ and 
$G_{\rm light}$), and hence $N^2(2N^2+1)\sim 2N^4$ correlation functions, so 
attempting to fit more than 2 modes is impractical.  We thus consider 
``$2\times 2\times 2$'' gravity: high/low redshift, large/small scales, 
$G_{\rm matter}$ and $G_{\rm light}$ \cite{danlin}. 

\begin{figure}[!hbtp] 
\begin{center}
\includegraphics[width=0.49\textwidth]{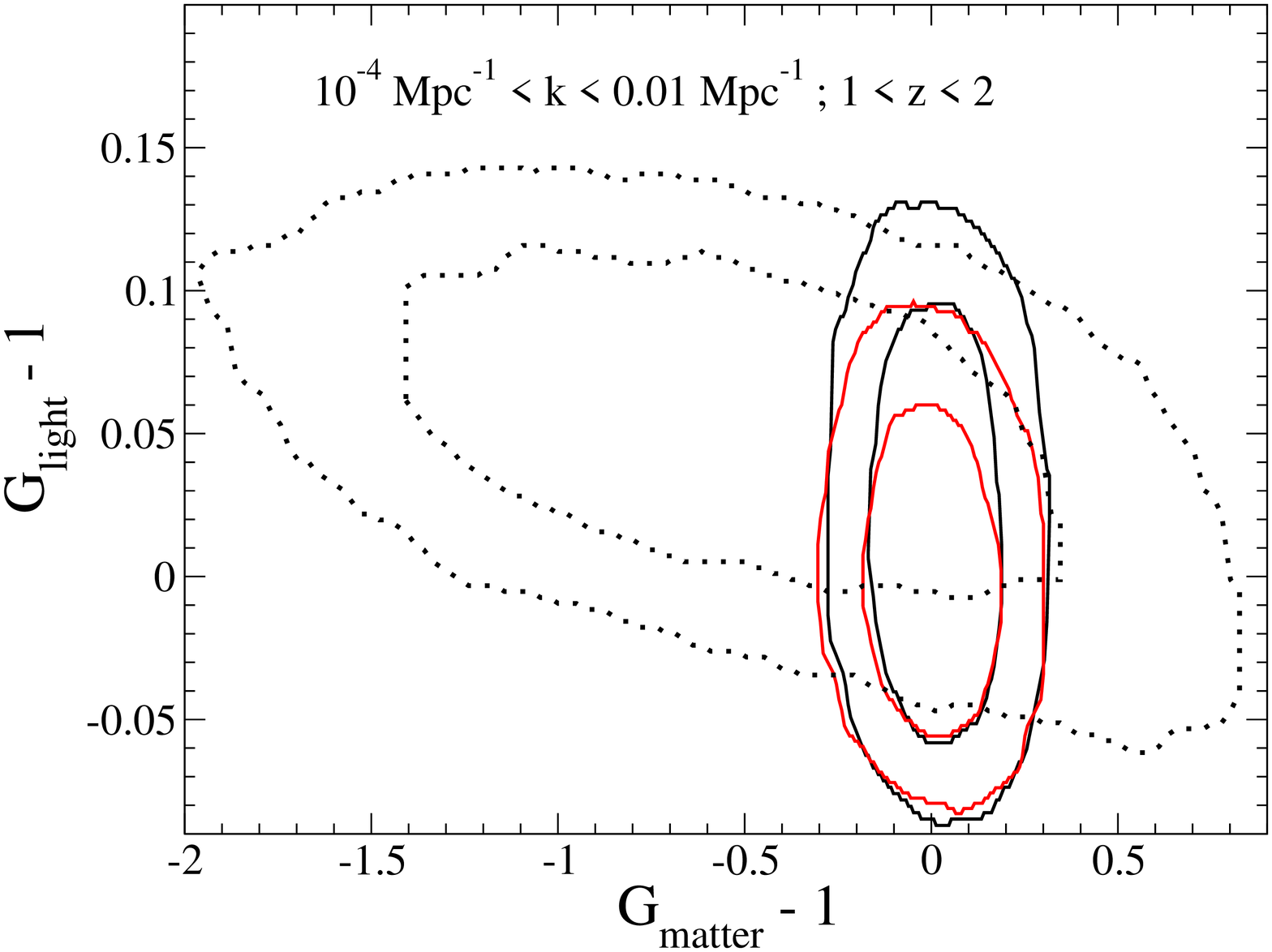}
\includegraphics[width=0.49\textwidth]{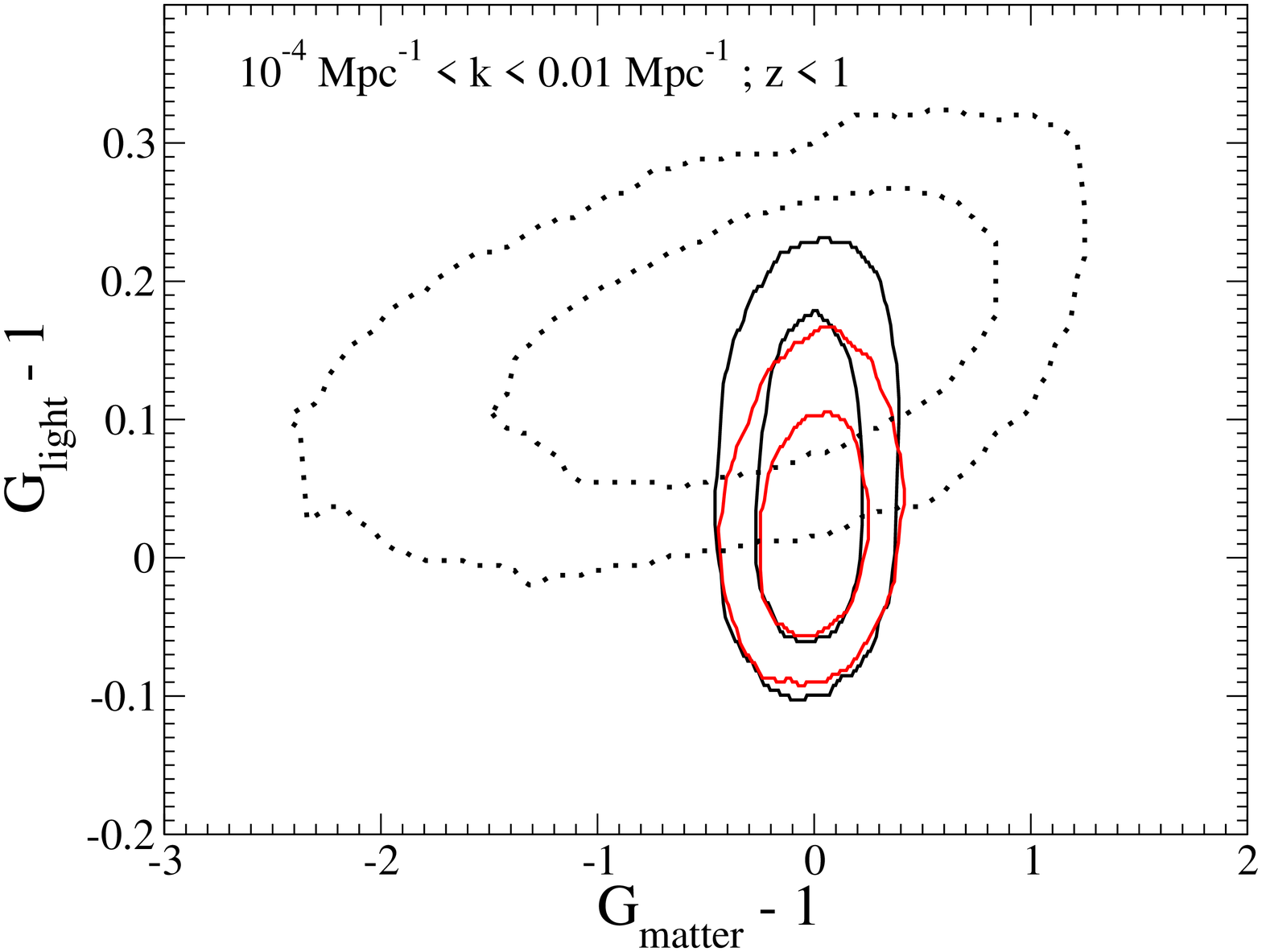}\\ 
\includegraphics[width=0.49\textwidth]{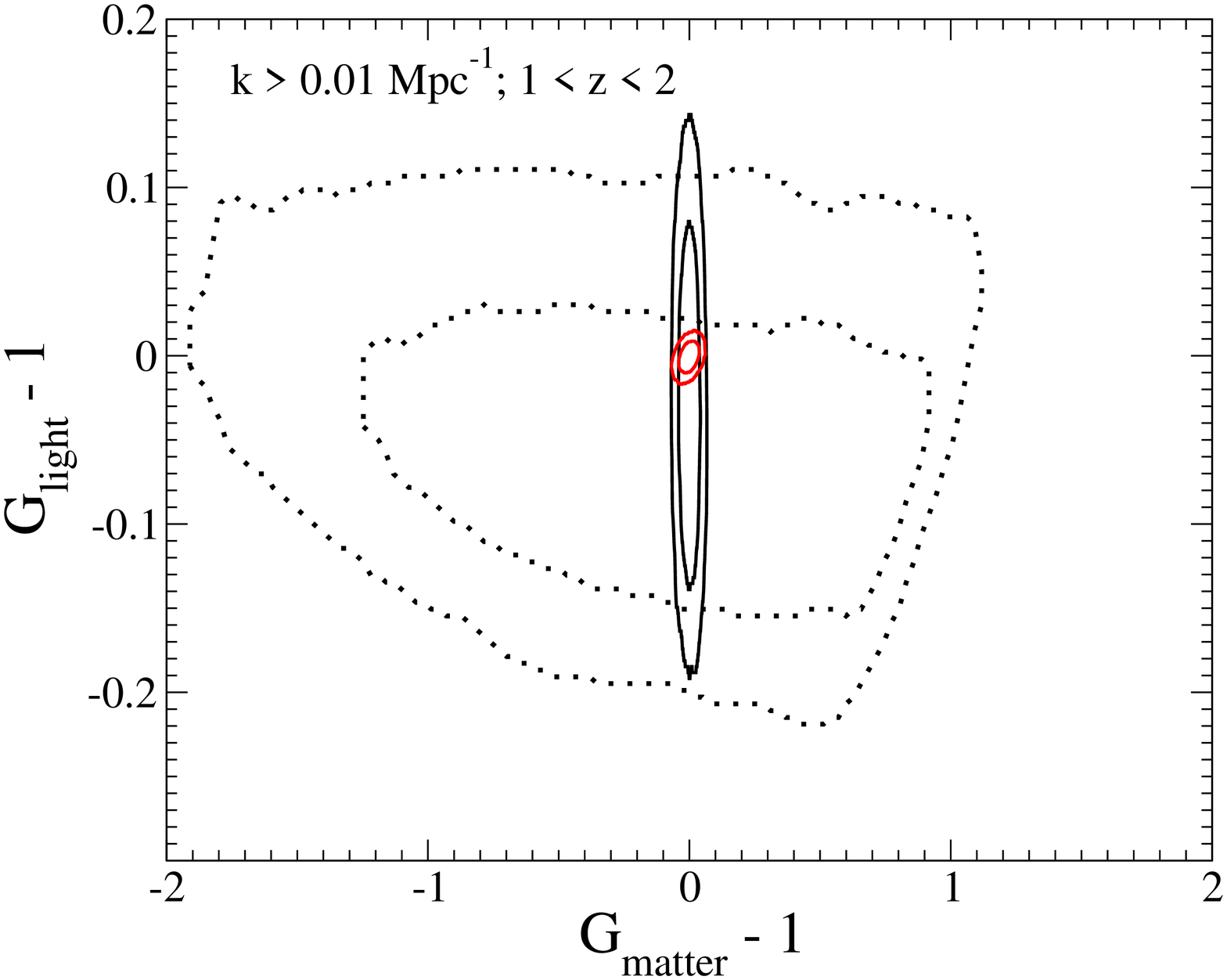}
\includegraphics[width=0.49\textwidth]{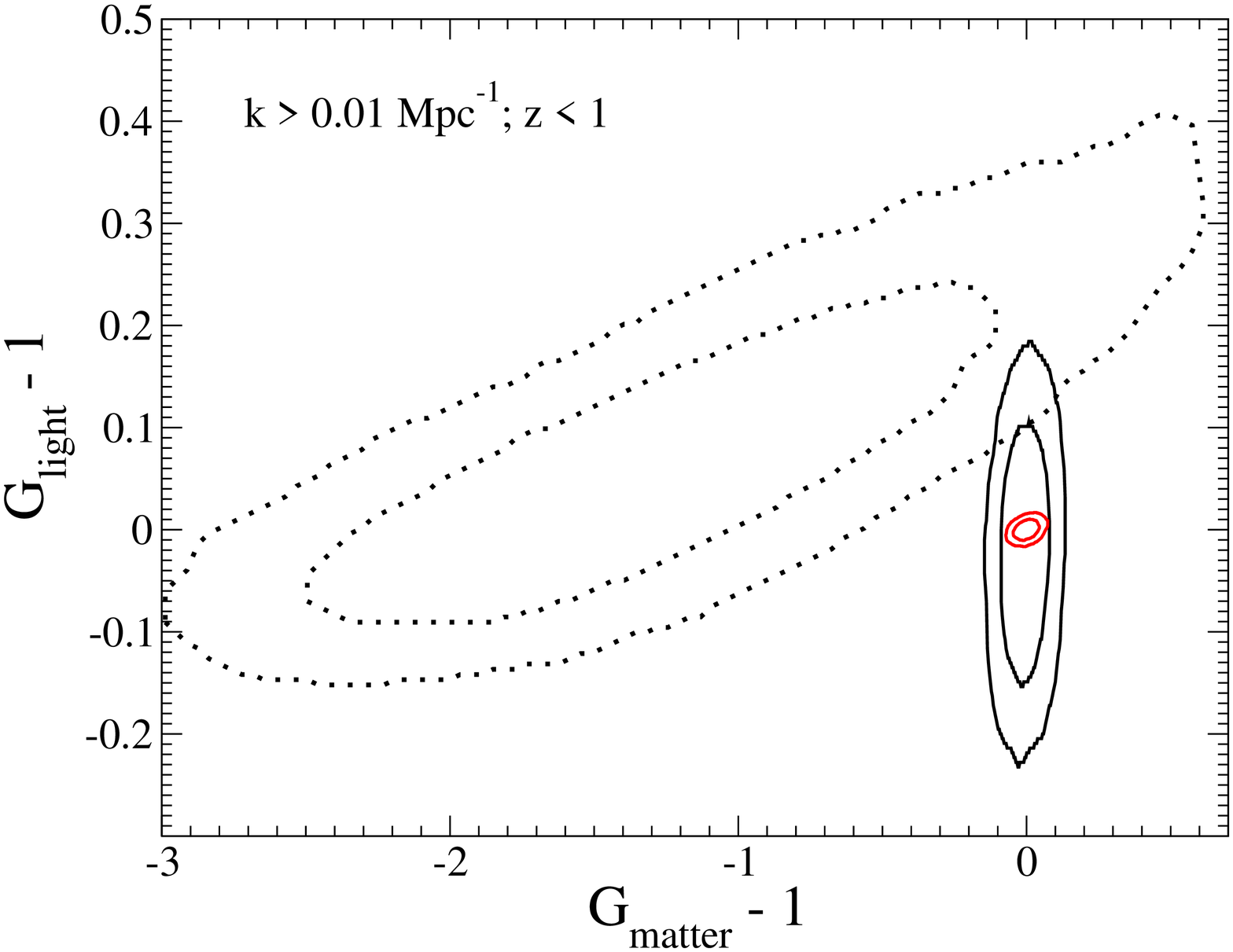}
\end{center}
\caption{\label{fig:gpanels}Constraints are shown on the 8 parameters 
of post-GR model independent gravity, from current and simulated future 
data.  Dotted contours show the 68\% and 95\% confidence regions from 
current WMAP CMB \cite{komatsu}, CFHTLS lensing \cite{fu}, and Union2.1 
supernova distance data \cite{suzuki}, 
while solid black contours show constraints from simulated data of the 
next generation BigBOSS galaxy redshift survey (marginalized over galaxy 
bias) plus Stage III experiments.  Small red contours include as well 
a highly optimistic galaxy weak lensing survey covering 4000 deg$^2$ 
with a galaxy number density of 55 arcmin$^{-2}$. 
}
\end{figure}

Such an approach is fully model independent and we can investigate the 
leverage of current and future observations to test cosmic gravity. 
In Figure~\ref{fig:gpanels} we show the results of Markov Chain Monte 
Carlo fits to data and simulations in the two bins each of wavenumber and 
redshift.  Galaxy redshift surveys such as BigBOSS \cite{bigboss} will 
be powerful ``gravity machines'', capable of mapping the density and 
velocity fields over a wide range of redshifts.  This is particularly 
useful at constraining $G_{\rm matter}$, the function sensitive to growth. 
Gravitational lensing information and the integrated Sachs-Wolfe effect 
of the cosmic microwave background act to constrain $G_{\rm light}$.

We see that current data are weak in the ability to test $G_{\rm matter}$, 
with uncertainties of order one, while $G_{\rm light}$ can be determined to 
roughly 10\% from the Canada-France-Hawaii Telescope Legacy Survey lensing 
data \cite{fu} 
and the WMAP CMB temperature power spectrum \cite{komatsu}.  However, note 
that this precision does not indicate accuracy; the deviation from GR seen 
in the bottom right panel, for example, arises from acknowledged observational 
systematics \cite{danlin}.  Galaxy redshift surveys such as BigBOSS will 
bring $G_{\rm matter}$ also to the 10\% precision level.  Recall that this 
is just what the ``paths of gravity'' phase space approach above argued was 
necessary to test GR, so we anticipate that next generation observations 
will reach interesting leverage in understanding cosmic gravity. 

For the galaxy surveys one must marginalize over astrophysical effects 
such as galaxy bias.  Here we have taken care of that by allowing the 
bias to float freely in each 0.1 bin of redshift.  Due to the 
complementarity with the other techniques included in Stage III experiments, 
this marginalization causes little degradation ($\lesssim10\%$) in 
the final constraints.  To reduce the uncertainties on $G_{\rm matter}$ 
further, one would need to increase the survey volume or use other growth 
probes such as CMB lensing.  To show the effect of a next generation 
galaxy weak lensing survey on $G_{\rm light}$, we include a rather 
optimistic survey in the inner red contours, with the result that 
$G_{\rm light}$ could be determined in some bins at the 1\% level.  In 
any case, the future of testing gravity looks extremely promising.

%%%%%%%%%%%%%%%%%%%%%%%%%%%%%%%%%%%%%%%%%%%%%%%%%%%%%%%%%%%%%%%%%%%%%% 
\section{The Paths of Gravity \label{sec:gal}} 

Exploring beyond the simplest models of gravity is useful to understand 
what evolutionary signatures the gravitational modifications might 
exhibit.  This allows us to investigate how robust are the well defined 
phase space tracks predicted by those particular models, and to look for 
characteristics that might be smoothed over by the model independent 
binned approach of the previous section. 

Moreover, $f(R)$ models are arbitrary and unnatural in the same sense that 
a quintessence potential $V(\phi)$ is: we have no clear physics guidance as 
to the appropriate functional form and high energy physics corrections will 
in any case alter it.  Two ways to protect against these problems 
are to avoid completely any potential, using only kinetic terms, and to 
look for a theory where the degrees of freedom arise geometrically, e.g.\ 
from higher dimensions, protecting against corrections.  Galileon gravity 
incorporates both these solutions.  

Galileon gravity is based on a shift symmetry in the effective field $\pi$ and 
involves nonlinear kinetic terms that are allowed (and only these are 
allowed) in order to guarantee second order field equations 
\cite{nicolis,deffayet}.  These also provide a Vainshtein screening that 
restores the theory to general relativity on small scales.  Thus, this 
is a well defined, robust theory that we can explore as a possible 
extension to GR. 

We can write the action for the Galileon theory as 
\begin{eqnarray} 
S &=& \int d^4 x\,\sqrt{-g} \left[ \left(1-2c_0\frac{\pi}{M_{\rm pl}}\right) 
\frac{M_{\rm pl}^{2} R}{2} - \frac{c_{2}}{2} (\partial \pi)^{2} 
 - \frac{c_{3}}{M^{3}}(\partial \pi)^{2} \Box \pi - 
\frac{c_{4}{\cal L}_{4}}{2} - \frac{c_{5}{\cal L}_{5}}{2} \right. \nonumber \\ 
&\qquad& \left. - \frac{M_{\rm pl}}{M^{3}} c_{\rm G} G^{\mu\nu} \partial_{\mu}\pi\partial_{\nu}\pi  - {\cal L}_{\rm m}    \right] \ . 
\end{eqnarray} 
General relativity arises simply from the ``1'' inside the square brackets. 
The terms involving $c_2$ through $c_5$ comprise the standard Galileon 
extension \cite{nicolis,deffayet}, and one can consider additional couplings 
to matter given by the $c_0$ term for a linear coupling and $c_G$ for a 
derivative coupling.  
Again, these terms guarantee second order field equations. 

From the action one can derive the background equations of motion and find 
under what circumstances one gets late time acceleration.  Note that 
acceleration occurs despite the absence of any cosmological constant or 
indeed any scalar field potential.  Galileon cosmology has attractor 
solutions in the radiation and matter eras, avoiding fine tuned sensitivities 
to initial conditions.  The effective equation of state $w$ at early times 
can be close to that of matter, allowing for a nonnegligible contribution 
of early dark energy density.  The field generically becomes phantom, 
$w<-1$, at some epoch near today, and has a future attractor to a de Sitter 
state.  These are all interesting properties for a theory to possess, and 
are discussed in some detail in \cite{applin}. 

Perturbing the equations of motion leads to the modified Poisson equations 
for $G_{\rm matter}$ and $G_{\rm light}$, and the growth of structure. 
We can then study the paths of gravity, the evolution of these $G(a)$ 
functions.  In the early universe Galileon gravity acts as a thawing 
field, locked to general relativity at high redshift and then gradually 
deviating.  The deviation increases linearly as the effective dark 
energy density does: $G\sim 1+b\,\Omega_\pi$.  However, unlike the DGP 
and $f(R)$ cases, the evolution is not a simple, quasi-parabolic 
trajectory from GR to a frozen new value of the gravitational strength. 
The interaction between the multiple Galileon terms leads to nonmonotonic 
behavior, strengthening gravity at redshift $z\sim 10$ then restoring 
to GR before deviating again toward a de Sitter attractor.  Interestingly, 
in this asymptotic state the gravitational slip, the difference between 
the metric potentials $\phi$ and $\psi$ with which we started our exploration 
of extensions to GR, vanishes (and hence $G_{\rm matter}=G_{\rm light}$). 
But although $\phi=\psi$ this is not general relativity, i.e.\ the 
strength of gravity differs from Newton's constant, $G\ne 1$.  

\begin{figure}[!hbtp]
\begin{center}
\includegraphics[angle=-90,width=\textwidth]{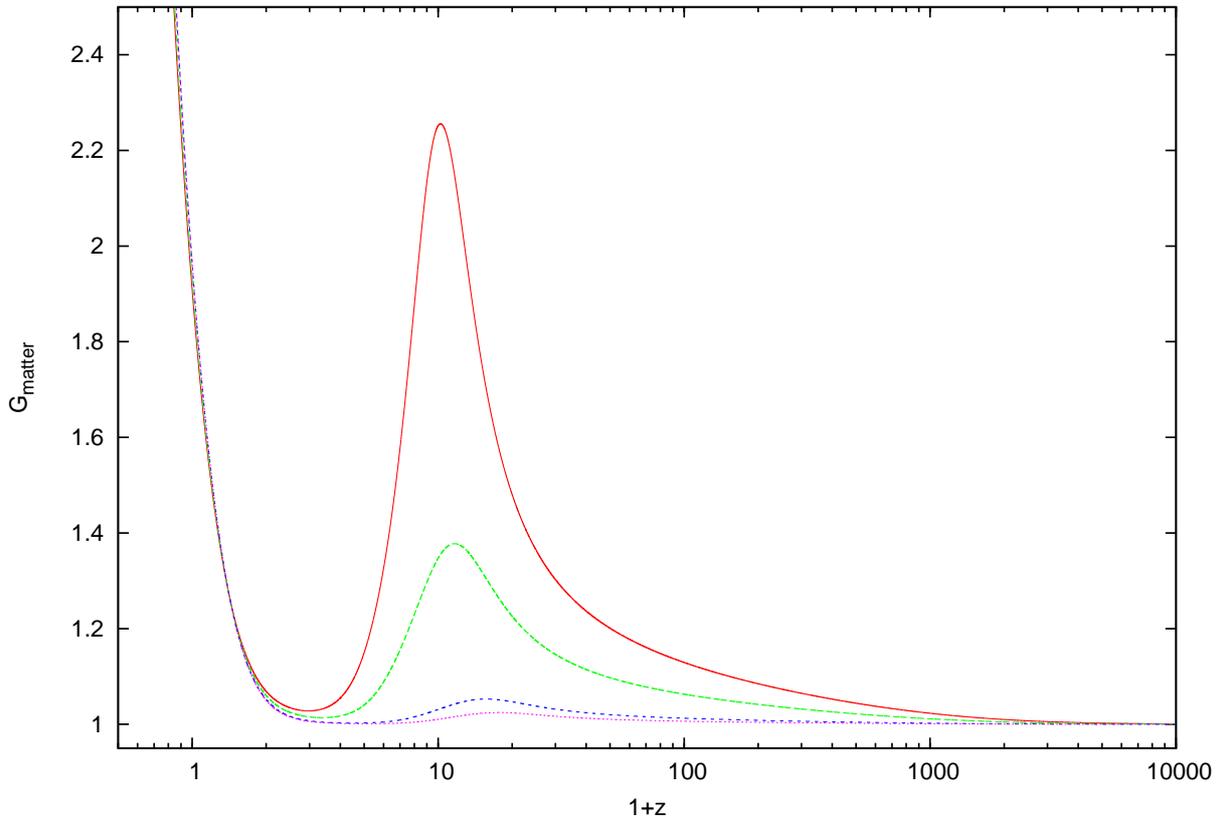} 
\end{center}
\caption{\label{fig:varyrhoi} The gravitational strength evolution 
varies with the initial dark energy density.  Here the curves have 
$\rho_{de}(z=10^6)=10^{-4}$, $5\times 10^{-5}$, $10^{-5}$, and 
$5\times 10^{-6}$ from top to bottom.  While the bump in strength 
grows and shifts to later times for higher density, the early time 
and late time behaviors are on attractors.  An interesting complementarity 
exists with the effective dark energy equation of state such that the less 
deviation from GR in $G_{\rm matter}$, the stronger phantom deviation from 
$w=-1$. 
} 
\end{figure}

Galileon cosmology thus shows some fascinating properties and signatures 
by which we can hope to distinguish it observationally from GR.  
Figure~\ref{fig:varyrhoi} shows the paths of gravity, the evolution of 
$G_{\rm matter}(a)$ and $G_{\rm light}(a)$ for the uncoupled Galileon 
for different values of the initial dark energy density.  The larger 
the initial density, the stronger the deviation from GR at $z\sim 10$ and 
the closer to the present the bump occurs; a low initial density is still 
capable of giving the same late time behavior but with an earlier and milder 
first deviation from GR.  One might conjecture whether the early 
enhancement in the gravitational strength could play a role in the 
development of high redshift structure and early massive galaxy clusters.

Detailed comparison of the predictions of Galileon cosmology for the 
expansion and growth history with observational data is in progress 
\cite{applin2}.  However, physical soundness of the theory in terms of 
lacking pathologies or instabilities already puts important constraints 
on the Galileon parameters \cite{applin}.  The most important conditions are 
the no-ghost condition, preventing the energy from being unbounded from below, 
and stability of perturbations, which can be written in terms of the 
sound speed as $c_s^2\ge 0$.  These requirements not only restrict the 
viable parameter space but in the linear and derivative 
coupling cases force those contributions to be subdominant to the standard 
Galileon terms at high redshift.  

Galileon cosmology has a much richer phenomenology than the simple 
extensions to GR previously considered and we are still exploring its 
implications.  It has a sounder theoretical foundation than many other 
theories (including most quintessence models) and the combination of 
theoretical and observational constraints may soon give rise to definite 
predictions for how to find deviations from general relativity.

%%%%%%%%%%%%%%%%%%%%%%%%%%%%%%%%%%%%%%%%%%%%%%%%%%%%%%%%%%%%%% 
\section{Conclusions} 

Gravitation is a ubiquitous and fundamental force but one that has 
not yet been rigorously tested, especially over the immense extrapolation 
to cosmic scales.  The ``direction of gravity'' is an open, active area 
of research and may hold key insights into cosmic acceleration and the 
growth of large scale structure. 

Measuring the expansion history alone, i.e.\ the dark energy equation of 
state  $w(a)$, is not sufficient to reveal the physics; rather, the 
combination of the expansion history and growth history, or testing 
gravity as well as the equation of state, is essential.  This introduces 
the functions $G_{\rm matter}$ and $G_{\rm light}$ from the modified 
Poisson equations, detailing gravity's direction of matter growth and 
of light deflection.  Failure to fit simultaneously for expansion and 
gravity runs the risk of substantially biasing both results. 

Fortunately, simultaneous fitting is straightforward (with proper 
definition of parameters to separate the physics) and does not 
substantially degrade the constraints.  Next generation observations 
will carry a wealth of information on the cosmic density and velocity 
fields, and enable model independent gravity constraints on 8 post-GR 
parameters at the 
10\% or better level.  For the simplest extensions to general relativity 
this is within the precision necessary to distinguish the correct theory. 

Galileon cosmology offers a more robust foundation with respect to 
physical naturalness.  It also provides a rich phenomenology with 
attractor solutions in the early and late universe and multiple 
signatures in both the expansion and gravity behavior.  The parameter 
space can be constrained through both theoretical considerations and 
comparison to observations.  Although the exploration of the paradoxical 
topic of the direction of gravity is still in early days, the results 
so far are exciting -- robust, testable alternatives to general relativity. 
The next generation of cosmology surveys and theoretical studies should 
finally test gravity diligently on cosmic scales and address the mystery 
of cosmic acceleration.

\ack 

I gratefully acknowledge my collaborators Scott Daniel and Stephen Appleby 
for Figures 1 and 2 respectively, and thank the ICGC 2011 organizers for 
the invitation to speak and the wonderful hospitality.  
This work has been supported in part by the Director, 
Office of Science, Office of High Energy Physics, of the U.S.\ Department 
of Energy under Contract No.\ DE-AC02-05CH11231 and by World Class 
University grant R32-2009-000-10130-0 through the National Research 
Foundation, Ministry of Education, Science and Technology of Korea.

\end{document}